\documentclass[preprint2]{aastex}

\usepackage{graphicx}
\usepackage{ulem}

\slugcomment{Not to appear in Nonlearned J., 45.}

\shorttitle{Arches cluster}
\shortauthors{Krivonos et al.}

\begin{document}


\title{First hard X-ray detection of the non-thermal emission around the Arches
  cluster: morphology and spectral studies with \textit{NuSTAR}}




\author{
Roman~A.~Krivonos\altaffilmark{1},
John~A.~Tomsick\altaffilmark{1},
Franz~E. Bauer\altaffilmark{2,3},
Frederick~K.~Baganoff\altaffilmark{4},
Nicolas~M.~Barriere\altaffilmark{1},
Arash~Bodaghee\altaffilmark{1},
Steven~E. Boggs\altaffilmark{1},
Finn~E.~Christensen\altaffilmark{5},
William~W.~Craig\altaffilmark{6,1},
Brian~W. Grefenstette\altaffilmark{7},
Charles~J.~Hailey\altaffilmark{8},
Fiona~A.~Harrison\altaffilmark{7},
JaeSub~Hong\altaffilmark{9},
Kristin~K.~Madsen\altaffilmark{7},
Kaya~Mori\altaffilmark{8},
Melania~Nynka\altaffilmark{8},
Daniel~Stern\altaffilmark{10},
William~W.~Zhang\altaffilmark{11}
}






\altaffiltext{1}{Space Science Lab, University of California, Berkeley, CA 94720; krivonos@ssl.berkeley.edu}
\altaffiltext{2}{Instituto de Astrof\'{\i}sica, Facultad de F\'{i}sica, Pontificia Universidad Catlica de Chile, 306, Santiago 22, Chile}
\altaffiltext{3}{Space Science Institute, 4750 Walnut Street, Suite 205, Boulder, Colorado 80301}
\altaffiltext{4}{MIT Kavli Institute for Astrophysics and Space Research, Cambridge, MA 02139, USA}
\altaffiltext{5}{DTU Space - National Space Institute, Technical University of Denmark, Elektrovej 327, 2800 Lyngby, Denmark}
\altaffiltext{6}{Lawrence Livermore National Laboratory, Livermore, CA 94550}
\altaffiltext{7}{Cahill Center for Astronomy and Astrophysics, California Institute of Technology, Pasadena, CA 91125}
\altaffiltext{8}{Columbia Astrophysics Laboratory, Columbia University, New York, NY 10027}
\altaffiltext{9}{Harvard-Smithsonian Center for Astrophysics, 60 Garden Street, Cambridge, MA 02138, USA}
 \altaffiltext{10}{Jet Propulsion Laboratory, California Institute of Technology, Pasadena, CA 91109}
 \altaffiltext{11}{NASA Goddard Space Flight Center, Greenbelt, MD 20771}



\begin{abstract}
  The Arches cluster is a young, densely packed massive star cluster
  in our Galaxy that shows a high level of star formation
  activity. The nature of the extended non-thermal X-ray emission
  around the cluster remains unclear. The observed bright Fe~K$\alpha$
  line emission at $6.4$~keV from material that is neutral or in a low
  ionization state can be produced either by X-ray photoionization or
  by cosmic-ray particle bombardment or both. In this paper we report
  on the first detection of the extended emission around the Arches
  cluster above 10~keV with the \textit{NuSTAR} mission, and present
  results on its morphology and spectrum. The spatial distribution of
  the hard X-ray emission is found to be consistent with the broad
  region around the cluster where the $6.4$~keV line is observed. The
  interpretation of the hard X-ray emission within the context of the
  X-ray reflection model puts a strong constraint on the luminosity of
  the possible illuminating hard X-ray source. The properties of the
  observed emission are also in broad agreement with the low-energy
  cosmic-ray proton excitation scenario.
\end{abstract}





\section{Introduction}

The Arches cluster \citep[G0.12+0.02,][]{cotera1996,serabyn1998} is a
massive star cluster with a core that is about $9''$ ($\sim0.35$~pc at
8~kpc) in radius \citep{figer1999}, with an average mass density of
$\sim 3\times 10^5 M_{\sun}\textrm{pc}^{-3}$, containing more than 160
O-type stars \citep{figer2002}. The Arches cluster is located in the
inner Galactic Center (GC) region at the projected angular distance of
$11'$ from the dynamic center of the Galaxy -- Sagittarius A*
(Sgr~A*).

The first serendipitous \textit{Chandra} observations of the Arches
cluster region revealed two bright sources (A1 and A2) located in the
cluster's core which are surrounded by the cluster's diffuse X-ray
emission \citep{zadeh2002}. A1 was later resolved into two distinct
bright sources: A1N and A1S by \cite{law2004} and \cite{wang2006}.
Subsequent dedicated campaigns revealed a complicated picture of the
Arches cluster X-ray emission, using spectral and morphological
studies to establish the presence of thermal and non-thermal emission
components.  The thermal emission is thought to originate from
multiple collisions between strong winds of massive stars
\citep{chlebowski1991,zadeh2002,wang2006,capelli11a}.  Diffuse
non-thermal X-ray emission has been detected from a broad region
around the cluster \citep{wang2006,tsujimoto2007,capelli11b,T12}. The
non-thermal nature of this extended radiation is revealed by its
bright fluorescent Fe~K$\alpha$~$6.4$~keV line emission. The
Fe~K$\alpha$ line at $6.4$~keV, which is due to Fe that is neutral or
in a low ionization state, is produced by ejecting a K-shell electron,
either by hard X-ray photoionization from an external X-ray source, or
by the collisional ionization induced by cosmic-ray (CR) accelerated
electrons or ions (mainly protons and $\alpha$-particles).

The mechanism producing the fluorescent line emission around the Arches
cluster is not completely understood. The fluorescence may be the
result of irradiation of the cloud by hard X-ray photons with energies
above $7.1$~keV (the K-edge of neutral iron). The required X-ray
illumination might be associated with a nearby transient X-ray source
\citep[e.g.][for 1E1740.7-2942]{churazov1993_1e1740} or with past
activity of Sgr~A$^*$, as suggested by \cite{sunyaev1993} to explain
the fluorescent line emission of the giant molecular cloud Sgr~B2 in
the GC region
\citep{koyama1996,sunyaev1998,murakami2000,revnivtsev2004a,terrier2010}. Alternatively,
the excitation of cold material might be through the collisional
ionization by CR particles
\citep[e.g.][]{predehl2003,zadeh2007,dogiel2009}. \cite{T12}
(hereafter T12) argued that bright 6.4 keV line structures observed
around the Arches cluster are very likely produced by bombardment of
molecular gas by energetic ions, and they disfavored the possibility
of accelerated electrons. T12 suggest that the required large flux of
low-energy cosmic ray ions could be produced in the ongoing supersonic
collision between the star cluster and a nearby molecular cloud.

Previously, the lack of spectral and morphological studies above 10
keV precluded us from building a complete picture of the Arches
cluster non-thermal emission and from making a definitive conclusion
about the ionizing mechanism. With the first focused observations of
the Arches cluster in the hard X-ray domain, we can shed light on the
cause of the non-thermal X-ray emission detected around the cluster.

We assume a distance of 8~kpc to the GC \citep{gillessen2009} and the
Arches cluster.  Sect.~\ref{data} of the paper describes the
\textit{NuSTAR} observations and data reduction. In
Sect.~\ref{morpho}, we study the morphology of the Arches cluster
using 2D image fitting procedures and wavelet image decomposition
techniques. In Sect.~\ref{spectra}, we present a spectral analysis
using the background approach specially developed for {\it NuSTAR}
observations. The main results are discussed in Sect.~\ref{discussion}
and summarized in Sect.~\ref{conclusion}.

\section{Observations and data reduction}
\label{data}

We observed Arches cluster during the GC region campaign with the {\it
  Nuclear Spectroscopic Telescope Array (NuSTAR)} \citep{nustar} in
October 2012. \textit{NuSTAR} operates in wide energy band from 3 to
79 keV, carrying two identical co-aligned X-ray telescopes with an
angular resolution of 18'' (FWHM). The focal planes of each telescope,
referred to as focal plane module A and B (FPMA and FPMB), provide a
spectral resolution of $400$~eV (FWHM) at $10$~keV.


\begin{table}
\begin{center}
\caption{Summary of the \textit{NuSTAR} observations available for the
  Arches cluster.\label{tab:data}}
\begin{tabular}{cccccc}
\tableline\tableline
Observation &  Date & \multicolumn{2}{c}{Exposure$^a$ (ks)} & Off-axis$^b$ \\
ID  &  2012   & FPMA  & FPMB  & (arcmin.) \\
\tableline
40010003001 &  Oct 14 & 24.2  & -- & 3.7\\
                     &             & (17.0) & -- & \\
40010005001 &  Oct 15 & 26.0 & 25.6 & 4.1 \\
                     &             & (15.7) & (20.0) &  \\
40010006001 & Oct 16 & 23.6  & 23.1 &  7.5 \\
                     &            & (12.0) & (12.1) &  \\
\tableline
\end{tabular}
\end{center}
Notes. $^a$Contains dead-time corrected nominal
  exposure, whereas the effective exposure, estimated for vignetting
  at 8~keV, is listed in parentheses.  
$^b$The off-axis angle is calculated for the Arches cluster
position with respect to the optical axis position on the detector.
\end{table}

The first part of the survey was carried out in the inner GC region
covering the area between Sgr A* and the persistent low-mass X-ray binary
1E~1743.1--2843 located $19'$ away
\citep{porquet2003a,delsanto2006}. The Arches cluster serendipitously
appears in the three \textit{NuSTAR} observations listed in
Table~\ref{tab:data}.

We processed the data from both modules using the \textit{NuSTAR Data
  Analysis Software (NuSTARDAS)} v1.2.0, the 2013 June 28 version
  of the \textit{NuSTAR} Calibration Database (CALDB), and HEASOFT
v6.13. The data were filtered for periods of Earth occultation, high
instrumental background due to South Atlantic Anomaly (SAA) passages,
and known bad/noisy detector pixels.

The \textit{NuSTAR} detectors are not entirely shielded from X-rays
that pass outside the X-ray optics modules and fall directly onto the
detectors.  Such an unfocused flux or ``stray-light'' can be
significant if there are bright X-ray sources within $2-3$ degrees of
the \textit{NuSTAR} field of view (FOV).  Stray-light from a point
source has a uniform distribution on the detector and can be easily
removed based on a geometrical model of the telescopes. We masked out
the stray-light region of the FPMB detector (FPMA is clean) caused by
two bright sources: GX~3+1 and the transient Swift~J174510.8-262411
\cite[see e.g.][and references herein]{vovk2012}, which was active at
the time of the \textit{NuSTAR} observations. The contaminated
detector pixels have been flagged as ``bad'' and removed from the
analysis at the stage of data screening in the \textit{NuSTARDAS}
pipeline run. The observation 40010003001 (FPMB) has been completely
disregarded because stray-light completely covered the Arches cluster
region.

\begin{figure*}[th]
\includegraphics[width=\textwidth]{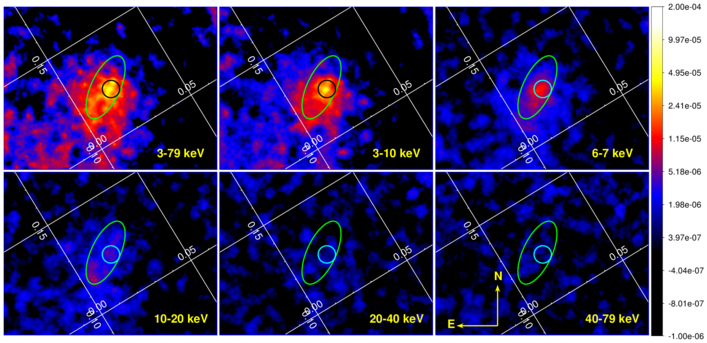}
\caption{\textit{NuSTAR} mosaic images of the Arches cluster in
  different energy bands. The images are background-subtracted, and
  adaptively smoothed with task \textit{dmimgadapt} from CIAO-4.5
  using a tophat kernel. The region used for local background
  estimation is shown in Figure~\ref{map:back}. The cyan circle (in
  black on the two upper images) denotes the Arches cluster emission
  prominent at $6.7$~keV (a He-like Fe~K$\alpha$ line), and this
  region is denoted as ``Cluster'' in Table~\ref{tab:regions}. The
  green ellipse represents the surrounding region characterized by the
  bright $6.4$~keV line emission (``Cloud'' region in
  Table~\ref{tab:regions}). Both regions have been adopted from T12
  for reference. The images are shown in the same log-scale color map
  shown on the right. The color bar is in units of
  cts~s$^{-1}$pix$^{-1}$, and the angular size of the image pixel
    is $2''$. The white grid indicates Galactic coordinates in
  degrees. The compass sign shows the image alignment in equatorial
  coordinates, with North up and East to the left.  The image
  orientation remains unchanged throughout the paper.}
\label{map:mosaics}
\end{figure*}

Stray-light photons from the extended sources and known celestial
X-ray backgrounds cannot be so easily removed because they fill the
whole detector area and have a non-uniform pattern depending on the
emission morphology. We treated this background differently for the
imaging and spectral analysis. Regarding the former, the non-uniform
background of the individual observations is effectively averaged when
it is combined into the mosaic, and therefore, can be considered as
nearly flat (however, see discussion in Sect.~\ref{core}). In
contrast, the non-uniformity of the background has a noticable effect
on the spectral analysis since it is done for individual
observations. To overcome this issue we applied the approach described
in Sect.~\ref{spectra}.

After running the pipeline and screening the data, we end up with
event lists for five \textit{NuSTAR} observations containing the
Arches cluster at different off-axis angles
(Table~\ref{tab:data}). Note that the source appears at large off-axis
distances, which causes the reduction of the efficiency (vignetting)
and strong PSF distortion. The total nominal (effective) dead-time
corrected exposure time of 122.5~(76.7)~ks comprises 73.8~(44.7)~ks
and 48.7~(32.1)~ks for FPMA and FPMB, respectively.

\section{Morphology}
\label{morpho}

The data from both detector modules were used to construct the mosaic
images (Figure~\ref{map:mosaics}) around the Arches cluster in
different energy bands. We used the $3-10$~keV \textit{NuSTAR} band as
a proxy for the standard $2-10$~keV energy range; $6-7$~keV band as
containing Fe~K$\alpha$ emission lines at $6.4$ and $6.7$~keV, which
are unresolved by \textit{NuSTAR}; and $10-20$~keV as a hard X-ray
band specific to \textit{NuSTAR}. The mosaics in the energy bands
$3-79$, $20-40$, and $40-79$~keV are also shown in
Figure~\ref{map:mosaics}. For each image an average (local) background
level was determined within an annular region centered on the Arches
cluster, as defined in Table~\ref{tab:regions} and illustrated in
Figure~\ref{map:back}. The background estimate was subsequently
subtracted from each individual image, prior to the construction of
the image mosaic. This is done because FPMB has a higher background
count rate than FPMA, requiring that we adjust the relative detector
background normalizations between modules. Note that the visible
gradient in Figure~\ref{map:back} is produced mainly by the emission
from the GC molecular clouds \citep[see e.g.][]{ponti2010}.

The exposure map, utilized for mosaic creation, has not been corrected
for vignetting effects. We checked that the vignetting correction does
not significally improve the results of the imaging
analysis. Hereafter, we assume no vignetting correction unless
otherwise stated.

The primary goal of the current paper is to study the non-thermal
emission surrounding the Arches cluster. The emission is revealed by
fluorescent Fe~K$\alpha$ 6.4~keV line, as has been observed by
\textit{XMM-Newton} and reported in the recent papers by
\cite{capelli11a} and T12. We adopted T12 sky regions describing the
core of the Arches cluster prominent in Fe~K$\alpha$ 6.7~keV line
(``Cluster'' region) and the surrounding region characterized by
the fluorescent 6.4~keV line (called ``Cloud'' region). The 6.7
keV line, attributed to the cluster core, originates from the hot,
thermally-ionized plasma, probably from the colliding winds of the
massive stars. The 6.4~keV line emission clearly extends beyond the
Arches cluster core, and comes from material in neutral or low
ionization states filling (in projection) an extended area represented
by an ellipse as illustrated in Figure~\ref{map:mosaics}. The
ellipse is oriented approximately in the North-South direction with
increased brightness to the South. The parameters of the background
and reference regions are listed in Table~\ref{tab:regions}.

\subsection{Cluster emission}
\label{core}

\begin{figure}
\begin{center}
\includegraphics[width=\columnwidth]{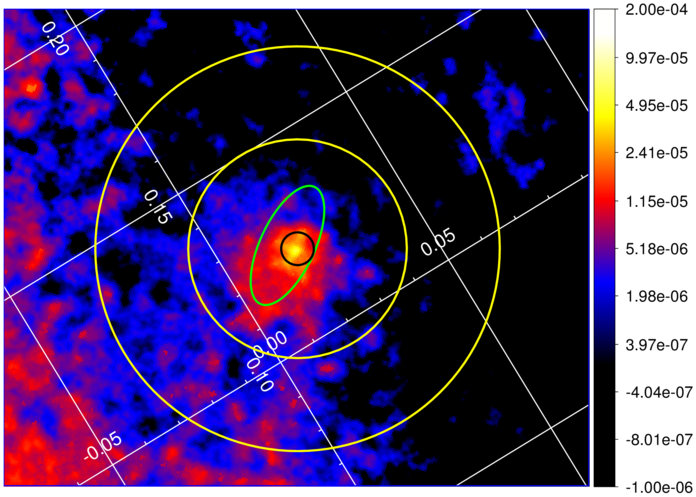}
\caption{The definition of the local background region around the
  Arches cluster. The image shows the \textit{NuSTAR} mosaic of the Arches
  cluster in the $3-10$~keV energy band (see Figure~\ref{map:mosaics} for
  the reference). The yellow annulus denotes the local background
  region used in the mosaic build up and fitting procedure
  (Sect.~\ref{core}). The parameters of the annulus are listed in
  Table~\ref{tab:regions}.}
\label{map:back}
\end{center}
\end{figure}

The Arches cluster has been relatively well studied in the standard
$2-10$~keV energy band with the \textit{Chandra} and
\textit{XMM-Newton} observatories.  Close attention has been paid to
the central part of the cluster, where thermal emission, thought to be
from colliding winds of giant stars, is observed. With data from
\textit{NuSTAR}, we can obtain new insight into the hard X-ray
emission of the Arches cluster.  However, even with angular resolution
that is unprecedented for hard X-ray astronomy ($18''$ FWHM), it is
challenging to resolve emission in the core, which has a radius of
$\sim15''$.

\begin{figure}
\begin{center}
\includegraphics[width=\columnwidth]{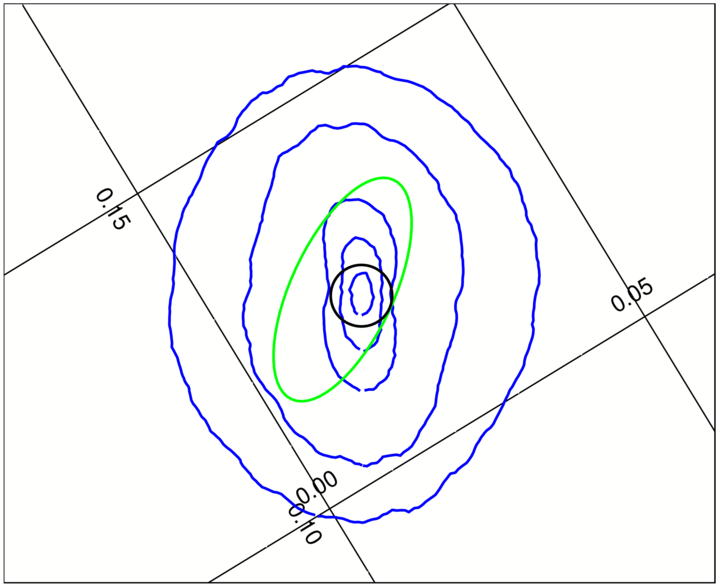}
\caption{The average \textit{NuSTAR} PSF in the $3-10$~keV energy
  range. The blue contours denote PSF encircled energy levels for 10,
  30, 50, 80, and 90\%, respectively, going from the center
  outwards. The PSF was averaged over five observations listed in
  Table~\ref{tab:data}. The green ellipse and black circle represent
  Arches cluster emission regions, see Figure~\ref{map:mosaics} for
  details.}
\label{fig:psf}
\end{center}
\end{figure}

The $3-10$~keV mosaic (Figure~\ref{map:mosaics}) demonstrates a strong
excess in the core of the Arches cluster and apparent extended
emission around it. To decouple these emission components and find the
centroid position of the core, we fitted a spatial model convolved
with an averaged PSF on the combined data set using the
\textit{Sherpa} package \citep{sherpa}, which is part of the
CIAO-4.5\footnote{http://cxc.cfa.harvard.edu/ciao} software
\citep{ciao}.  

\begin{figure}
\includegraphics[width=\columnwidth]{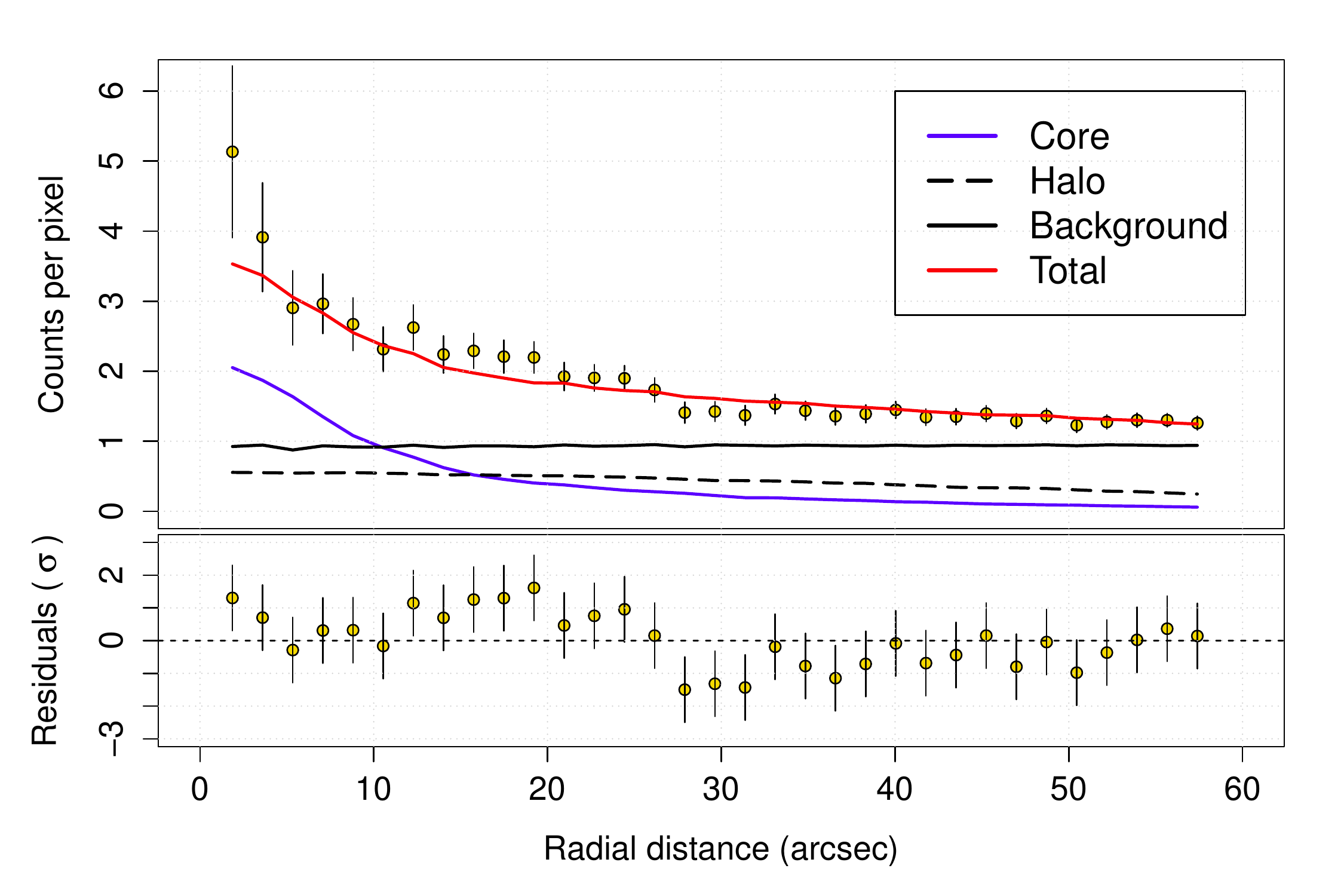}
\caption{Radial profile of the Arches cluster region in the $3-10$~keV
 energy band. Lines show the best-fit model and its components convolved
  with the averaged PSF.}
\label{fig:core}
\end{figure}

To make a PSF function representing the average over the different
source offsets in the ObsIDs, we used the inflight-calibrated PSF
shape stored in the CALDB. The PSF images were weighted according to
the exposure time of each data set. The resulting averaged PSF is
shown in Figure~\ref{fig:psf}. Its shape is strongly distorted because
of the large off-axis angles (on-axis PSF would be
symmetric). Importantly, the PSF elongation is not co-aligned with the
6.4~keV line cloud region. For the fitting procedure, we assume no
significant change of PSF shape inside the radius of $60''$ around the
Arches cluster (see below). Note that 90\% of the averaged PSF
encircled energy is situated at or inside of the inner radius of the
annular region used for the background estimation, depending on the
adopted energy used. Using this fact, we estimated the background
count rate in the separate fitting procedure without PSF convolution.

\begin{figure}
\begin{center}
\includegraphics[width=\columnwidth]{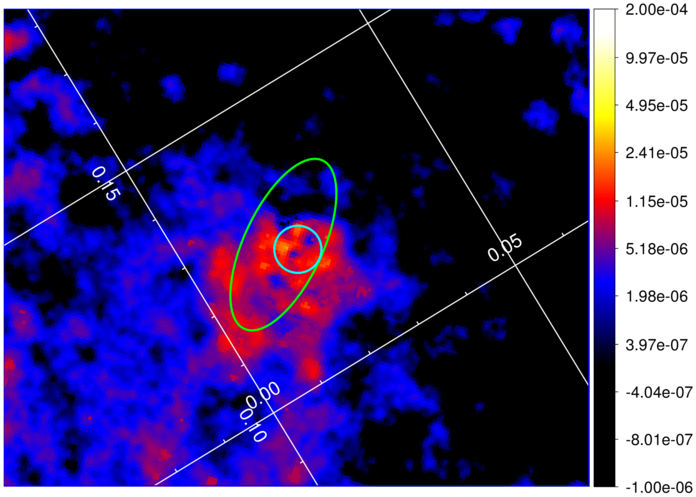}
\caption{The \textit{NuSTAR} $3-10$~keV residual image of the Arches
  cluster after removal of the emission from the cluster's core and
  background subtraction. The details of the region description can be
  found in Figure~\ref{map:mosaics}. The color bar units are
  cts~s$^{-1}$pix$^{-1}$ and the angular pixel size is $2''$.}
\label{map:resid}
\end{center}
\end{figure}

\begin{table}[htbp]
\begin{center}
  \caption{Definitions of the sky regions.\label{tab:regions}}
\begin{tabular}{cccccc}
\tableline\tableline
 RA (J2000) & Dec (J2000) & Parameters\\
\tableline
\multicolumn{3}{c}{Cluster (circle)} \\
$17^h45^m50.3^s$ & $-28^{\circ}49'19''$ & $15''$\\
\tableline
\multicolumn{3}{c}{Cloud (ellipse)} \\
$17^h45^m51.0^s$ & $-28^{\circ}49'16''$ & $25'',
59'', 155^{\circ}$ \\
$17^h45^m50.3^s$ & $-28^{\circ}49'19''$ & $15''$ (excl.)\\
\tableline
\multicolumn{3}{c}{Background (annulus)} \\
$17^h45^m50.3^s$ & $-28^{\circ}49'19''$ & $185''$\\
$17^h45^m50.3^s$ & $-28^{\circ}49'19''$ & $100''$ (excl.)\\
\tableline
\end{tabular}
\end{center}
Notes. Central position and radius for circular regions, and semi- minor/major axes and rotation angle for the
elliptical regions. The rotation is defined counter clockwise relative
to North (upward). ``Cluster'' and ``Cloud'' regions are illustrated
in Figure~\ref{map:mosaics}. The excluded regions are marked as ``excl.''.
\end{table}

Given the spatial extent of the emission around the Arches Cluster as
found by Chandra and XMM-Newton, we want to confirm whether we can
resolve this same emission with \textit{NuSTAR}. This will give us
confidence that we may also be able to spatially resolve the Arches
cluster above $10$~keV. 

We represented the Arches cluster core emission by a 2D Gaussian.  Due
to the low count rate per pixel, we used Poisson statistics in the
fitting procedure. The background term has been constrained in the
annulus region shown in Figure~\ref{map:back}.  After running the
fitting procedure with the 2D Gaussian model, we found that the
$3-10$~keV emission from the Arches cluster was not acceptably fitted
with a single point source, and an extended ``halo'' around the
cluster had to be considered. This finding is in line with the
detection of extended emission surrounding the Arches cluster as
observed by \textit{Chandra} and \textit{XMM-Newton}
\citep{zadeh2002,law2004,wang2006}.

\begin{figure*}
\includegraphics[width=0.49\textwidth]{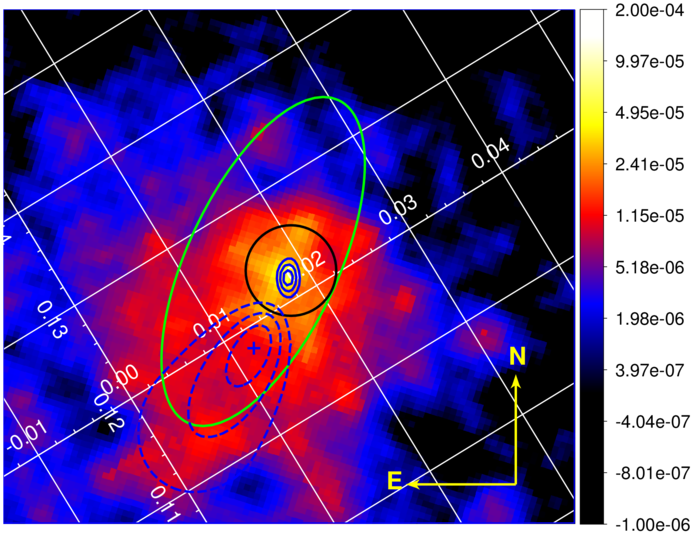}
\includegraphics[width=0.49\textwidth]{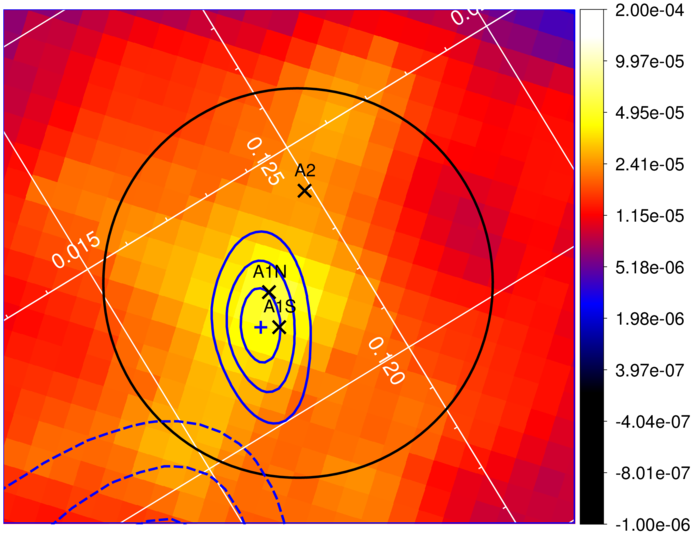}
\caption{{\it Left:} Arches cluster core and halo centroid
  determination on the $3-10$~keV \textit{NuSTAR} mosaic image (use
  Figure~\ref{map:mosaics} for the reference).  Blue contours show 1-,
  2- and 3-$\sigma$ confidence intervals for the core (solid blue) and
  the halo (dashed blue) centroids. The centroid positions of the core
  and halo emission are marked as blue crosses. {\it Right:} A
  close-up of the region of the Arches core. Black points mark
  positions of three bright sources A1S/N and A2 that are resolved by
  \textit{Chandra} observations of the Arches cluster
  \citep{wang2006}. Note that a systematic scatter of $0.9''$ (RMS) is
  not taken into account (see Sect.~\ref{core}). The color bar units
  are cts~s$^{-1}$pix$^{-1}$ and the pixel size is $2''$.}
\label{fig:centroid}
\end{figure*}

We constrained the fitting area to a circle centered on the core and
extending to a radius of $60''$. The model includes (i) a narrow
2D~Gaussian of fixed width ($4''$~FWHM, chosen to emulate the PSF
smearing effect caused by incomplete aspect reconstruction),
representing the cluster's core, (ii) a wide 2D~Gaussian describing
the halo, and (iii) a constant background term. Note that the visible
in Figure~\ref{map:back} background gradient has no significant effect
on the fitting procedure. The best-fit model parameters are listed in
Table~\ref{tab:pars}. Figure~\ref{fig:core} shows the radial profile
of the Arches cluster intensity distribution centered at the best-fit
position of the core.

To demonstrate a surface brightness distribution of the Arches cluster
halo, we show in Figure~\ref{map:resid} the residuals after
subtracting off a best-fit point source corresponding to the core
emission.

The position of the cluster's core is spatially consistent with the
brightest source, $A1$, detected by \textit{Chandra} in the Arches
cluster \citep{zadeh2002} and later resolved into the two sources
$A1S/N$ \citep{law2004,wang2006}. Figure~\ref{fig:centroid} shows
confidence regions for the position uncertainty of the core and halo
components.

Note that due to the \textit{NuSTAR} optics design, the celestial
coordinates of each incoming photon are determined during the post
processing. This task is complicated by distortions due to thermal
bending and external forces acting on the mast during orbit. The
systematic offset of the source position determination can be as high
as $10''$. However, the fact that the position of the Arches cluster
core emission detected with \textit{NuSTAR} is in agreement (within
the angular separation of $\sim1.5''$) with the catalogued coordinates
of the two bright \textit{Chandra} sources $A1S/N$, tells us that the
described systematic offsets are negligible for these observations.

Our ability to centroid a weak source with \textit{NuSTAR} is also subject to
uncertainties based on how positions are probabilistically assigned to
photons in the \textit{NuSTARDAS} pipeline and the finite spatial
resolution of the \textit{NuSTAR} pixels. To estimate any scatter introduced by
this process we ran the \textit{NuSTARDAS} pipeline 50 times, repeated
out analyses, and found that the sky position of the Gaussian core has
an RMS scatter of $0.9''$. We adopt this as the systematic error on
the cluster core position.

The width of the 2D Gaussian halo model component does not 
have strong physical meaning since this component is used specifically 
to describe local background of the core. However, we should note that
the center position of the halo component does not coincide with the
core centroid, and the trend suggests that the halo centroid falls
outside the Arches cluster (Figure~\ref{fig:centroid}).

\begin{table}[htbp]
\begin{center}
  \caption{Best-fit model parameters for the PSF fitting procedure of
    the $3-10$~keV Arches cluster core and halo emissions.\label{tab:pars}}
\begin{tabular}{cccccc}
\tableline\tableline
Parameter & Value \\
\tableline
\multicolumn{2}{c}{Background (constant)} \\
Norm. $(\times 10^{-5})$  & $1.18\pm0.01$ cnts\ s$^{-1}$ pix$^{-1}$\\
\tableline
\multicolumn{2}{c}{Core (2D Gaussian)} \\
Center R.A.$^a$ & $17^h50^m50.43^{s}$  \\
Center Dec.$^a$ &   $-28^{\circ}49'23.07''$ \\
FWHM & $4.0''$ (fixed) \\
Norm. $(\times 10^{-3})$ & $2.2\pm 0.3$ cnts\ s$^{-1}$ pix$^{-1}$\\
\tableline
\multicolumn{2}{c}{Halo (2D Gaussian)} \\
Center R.A.$^a$ & $17^h45^m50.62^{s}$  \\
Center Dec.$^a$ &  $-28^{\circ}49'47.17''$ \\
FWHM & $72''.4^{+6''.7}_{-4''.7}$\\
Norm. $(\times 10^{-5})$ & $1.77^{+0.45}_{-0.37}$ cnts\ s$^{-1}$  pix$^{-1}$\\
\tableline
Cash statistic   &  4661.91 \\
Degrees of freedom  & 2820 \\
\tableline
\end{tabular}
\end{center}
Notes. $^a$J2000 epoch, the $1\sigma$ positional uncertainty is less
than $2''$ and $6-8''$ for the core and halo components, respectively, see 2D
confidence regions in Figure~\ref{fig:centroid} for details. 
Model normalization factor is marked as ``Norm.''.
\end{table}


\subsection{Wavelet decomposition}
\label{atrous}

\setlength{\unitlength}{\textwidth}
\begin{figure*}[th]
\begin{picture}(1.0,0.70)
\put(0.04,-0.03){\includegraphics[width=\textwidth]{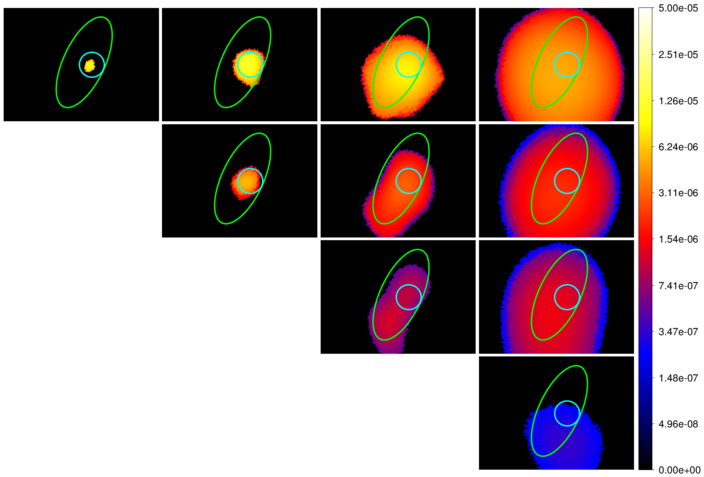} }
\put(0.1,0.645){$J=3$~$(8'')$}
\put(0.32,0.645){$J=4$~$(16'')$}
\put(0.55,0.645){$J=5$~$(32'')$}
\put(0.78,0.645){$J=6$~$(64'')$}
\put(0.02,0.51){\rotatebox{90}{$3-10$~keV}}
\put(0.24,0.36){\rotatebox{90}{$6-7$~keV}}
\put(0.47,0.18){\rotatebox{90}{$10-20$~keV}}
\put(0.69,0.01){\rotatebox{90}{$20-40$~keV}}
\end{picture}
\caption{Wavelet decomposition of \textit{NuSTAR} mosaic images of the
  Arches cluster in $3-10$, $6-7$, $10-20$, and $20-40$~keV energy
  bands. All structures with detection significance below $5\sigma$
  have been removed from the images. The units of the color bar on the
  right are cts~s$^{-1}$pix$^{-1}$, where the pixel size is $2''$.}
\label{map:scales1}
\end{figure*}
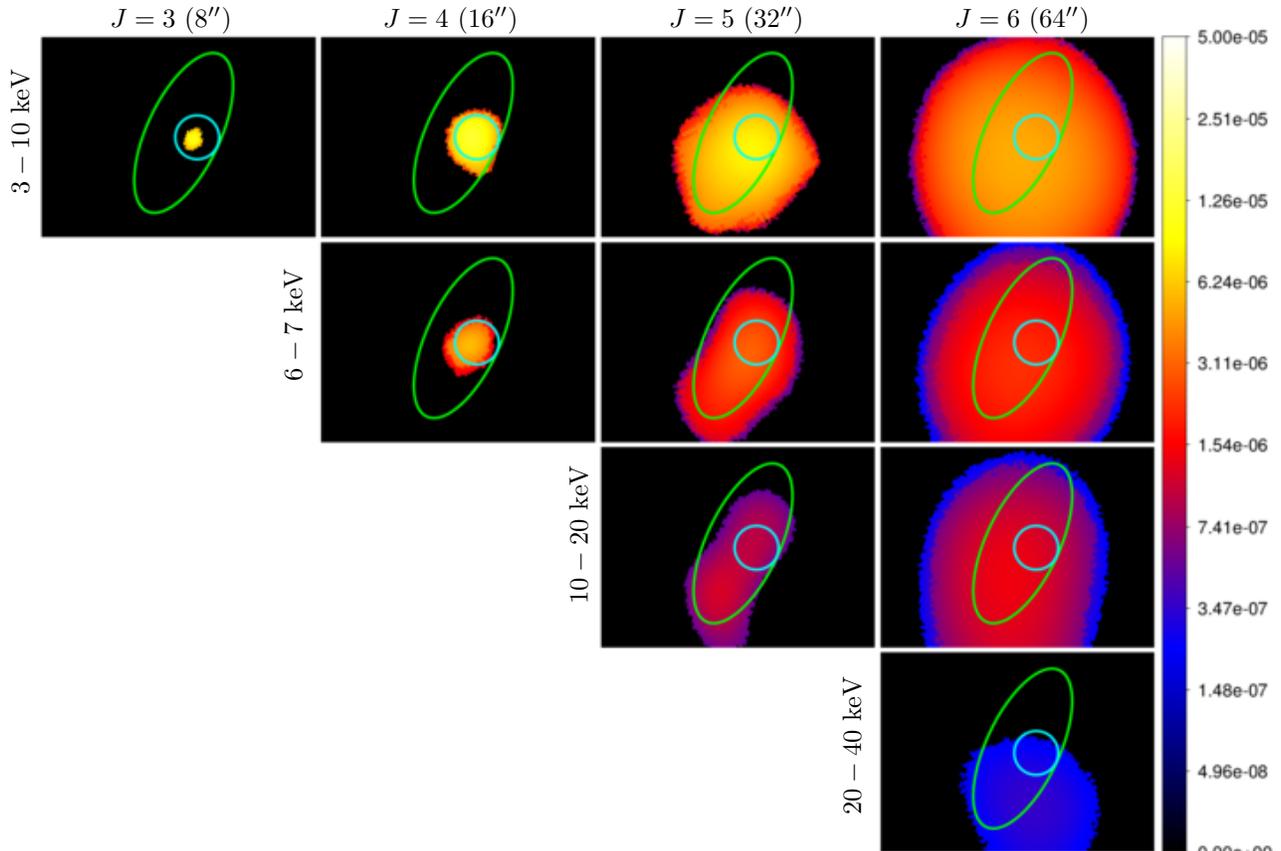

\begin{figure*}[th]
\includegraphics[width=0.34\textwidth]{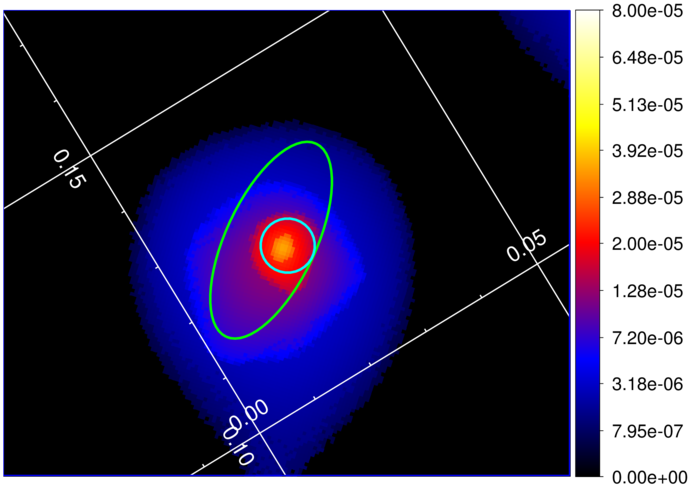}
\includegraphics[width=0.34\textwidth]{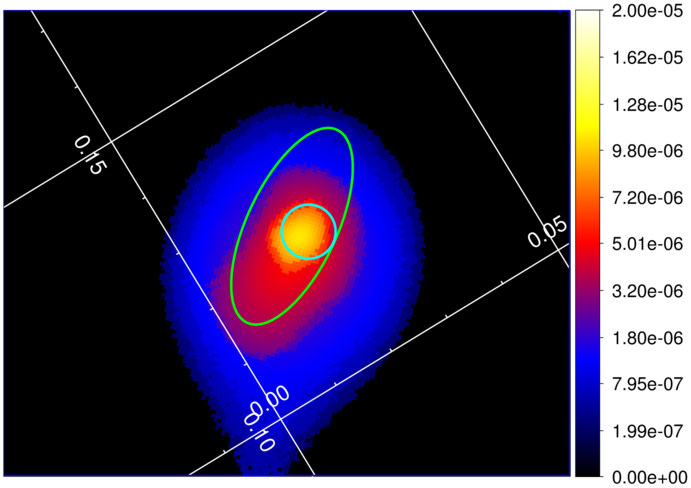}
\includegraphics[width=0.34\textwidth]{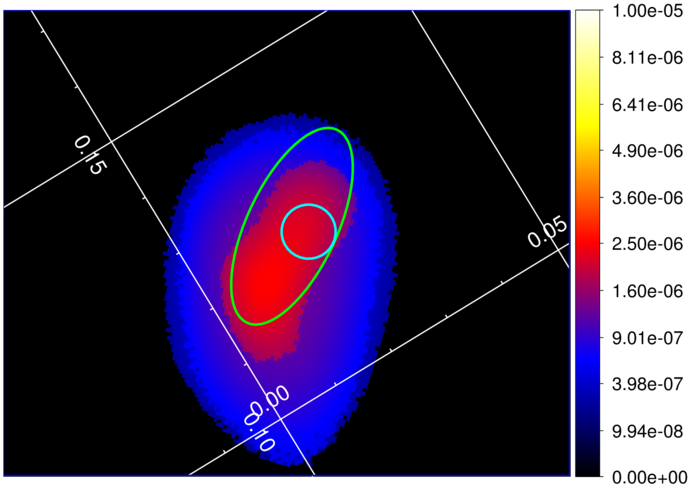}
\caption{Stacked wavelet layers (see Figure~\ref{map:scales1}) of the
  Arches cluster \textit{NuSTAR} mosaic images in the (from left to
  right) $3-10$, $6-7$, and $10-20$~keV energy bands. All structures
  with detection significance below $5\sigma$ have been removed from
  the images. The units of the images are cts~s$^{-1}$pix$^{-1}$
  and the pixel size is $2''$.}
\label{map:scales2}
\end{figure*}

We studied in Sect.~\ref{core} the morphology of the Arches cluster in the
$3-10$~keV band fitting data with a parametric model. We constrained 
the central part of the cluster, determined its centroid position, flux,
and revealed strong excess outside the cluster core. Since we observe
extended emission with a complicated projected morphology, it
becomes quite difficult to represent this emission with an analytical 
model, i.e. to parameterize it. In this case, \textit{non-parametric} 
methods can be used for morphological studies. One such method is a 
wavelet image transform.


For the decomposition of \textit{NuSTAR} images in different energy
bands, we used the so called \textit{\`a~trous}\/ digital wavelet
transform algorithm
\citep{starck94,slezak94,vikhlinin96,krivonos2010}. The main advantage
of this algorithm over other wavelet methods is the fact that it
decomposes the image into a linear combination of layers or
``scales'', each of them emphasizes the structures with a
characteristic size of $\approx 2^{J-1}$~pixels, which is
$2^{J}$~arcseconds in the case of our \textit{NuSTAR} mosaic image
resolution ($1\textrm{pix}=2''$). Thus, low (high) $J$ values
correspond to small (large) spatial variations. We used the {\it
  \small WVDECOMP} wavelet decomposition code developed by
\cite{wvdecomp,wvdecomp98} as a part of {\it \small ZHTOOLS} software
package\footnote{http://hea-www.harvard.edu/RD/zhtools}.

We ran {\it \small WVDECOMP} on mosaic images for scales $1-7$
($2''-128''$), detecting all significant structures above the
$5\sigma$ level. To keep the photon statistics on the images, we
provided counts and exposure maps to {\it \small WVDECOMP}. As
mentioned in Sect.~\ref{morpho}, FPMB has a higher background count
rate than FPMA. In fact, this can mimic extended structures on the
exposure-corrected image. To diminish this effect, we built the
background map for a given observation by multiplying the exposure map
and background count rate measured in the annulus region shown in
Figure~\ref{map:back}. The individual background maps were combined
into the mosaic and used as a part of the {\it \small WVDECOMP}
algorithm.

Figure~\ref{map:scales1} shows wavelet scales $3-6$ of the Arches
cluster mosaic images in $3-10$, $6-7$, and $10-20$~keV. The lowest
scales ($1,2$) do not contain significantly detected ($5\sigma$)
structures. The highest scale ($7$)
contains large-scale diffuse background attributed to the Galactic
center molecular complex zone \citep{ponti2010} and Galactic Ridge
X-ray Emission \citep[GRXE, see e.g.][]{revnivtsev2006}. We combined
scales $3-6$, which only contain structures related to the Arches
cluster into the mosaics shown in Figure~\ref{map:scales2}. The main
results from the WT decomposition can be summarized as follows (refer
to Figs.~\ref{map:scales1} and \ref{map:scales2}).

\uline{\it $3-10$~keV:} The core of the Arches cluster is clearly visible 
on the scales $J=3-4$ and the halo emission is revealed on scale
$J=5$. The apparent position of the halo is consistent with fitting
results in Sect.~\ref{core} showing a shifted position with respect
to the cluster core.


\uline{\it $6-7$~keV:} This energy band contains two blended
Fe~K$\alpha$ lines at $6.4$ and $6.7$ keV. We speculate that these
lines appear at scales $J=4$ (6.7~keV) and $J=5$ (6.4~keV). The
$6.7$~keV line is spatially consistent with core emission, while the
$6.4$~keV line emission follows the cloud region as expected. These
results are in full agreement with \textit{XMM-Newton} observations
(T12). In contrast to the $3-10$~keV band, the $6.7$~keV line emission
distribution does not demonstrate high-frequency detection at $J=3$,
which may indicate its broader spatial distribution in comparison with
the $3-10$~keV continuum emission of the Arches cluster. However we
should note that the $6-7$~keV mosaic image is not continuum
subtracted.

\uline{\it $10-20$~keV:} The Arches cluster does not show any
significant ($>5\sigma$) point-like structures in the core region,
which is expected due to the rapid fall of the $\sim 1.7$~keV thermal
emission above $10$~keV. On the other hand, the hard X-ray emission is
significantly detected at scales $J=5-6$. Note that the observed
extended structure cannot be caused by the wide wings of the PSF -- as
is demonstrated in Figure~\ref{fig:psf}, the PSF shape is not aligned
with the cloud region morphology. The observed $10-20$~keV continuum
emission apparently coincides with Fe~K$\alpha$ fluorescent $6.4$~keV
line emission observed with \textit{XMM-Newton} (T12). According to
T12, the $6.4$~keV line emission demonstrates brightening in the
south-east part of the ellipse, which is also noticeable at scale
$J=5$.

\uline{\it $20-40$~keV:} The mosaic map of the Arches cluster shown in
Figure~\ref{map:mosaics} does not allow to make a definitive conclusion
about the source detection. However, wavelet decomposition shows weak
excess at scale $J=6$. This fact is implicitly confirmed by detection
of the emission up to 40 keV in the spectral analysis of
Sect.~\ref{spectra}. The apparent hard X-ray emission is located outside the
cluster and consistent with general extent of $10-20$~keV emission.

\section{Spectroscopy} 
\label{spectra}

We extracted spectral information for the Arches cluster from the
circular region with radius of $50''$ centered at the cluster's core
(Table~\ref{tab:regions}), including the core and extended
emission. The spectra have been extracted from all observations using
the \textit{nuproducts} task applying PSF and ghost ray corrections to
the effective area ARF (Ancillary Response File).

\begin{figure*}[ht]
\includegraphics[width=0.34\textwidth]{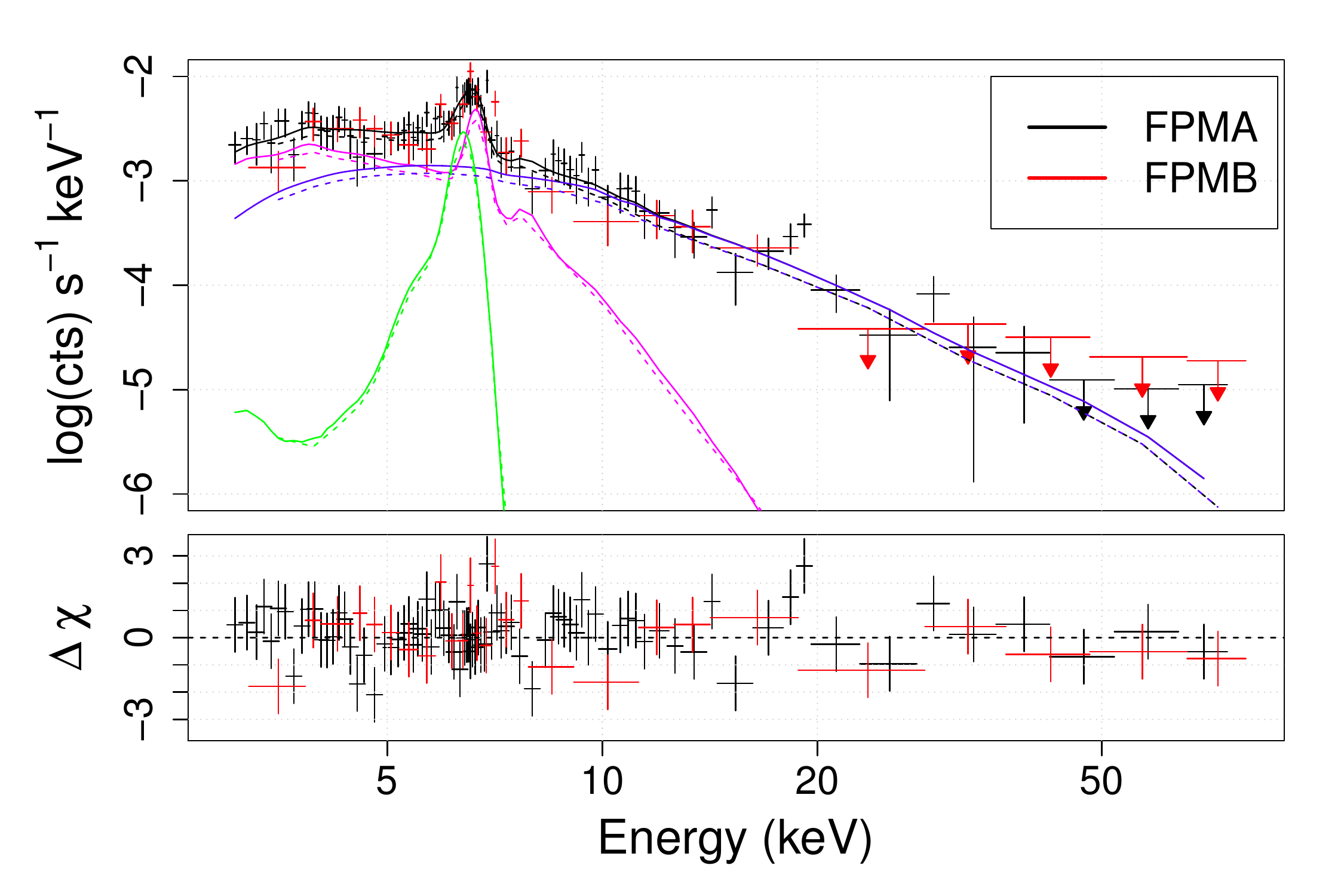}
\includegraphics[width=0.34\textwidth]{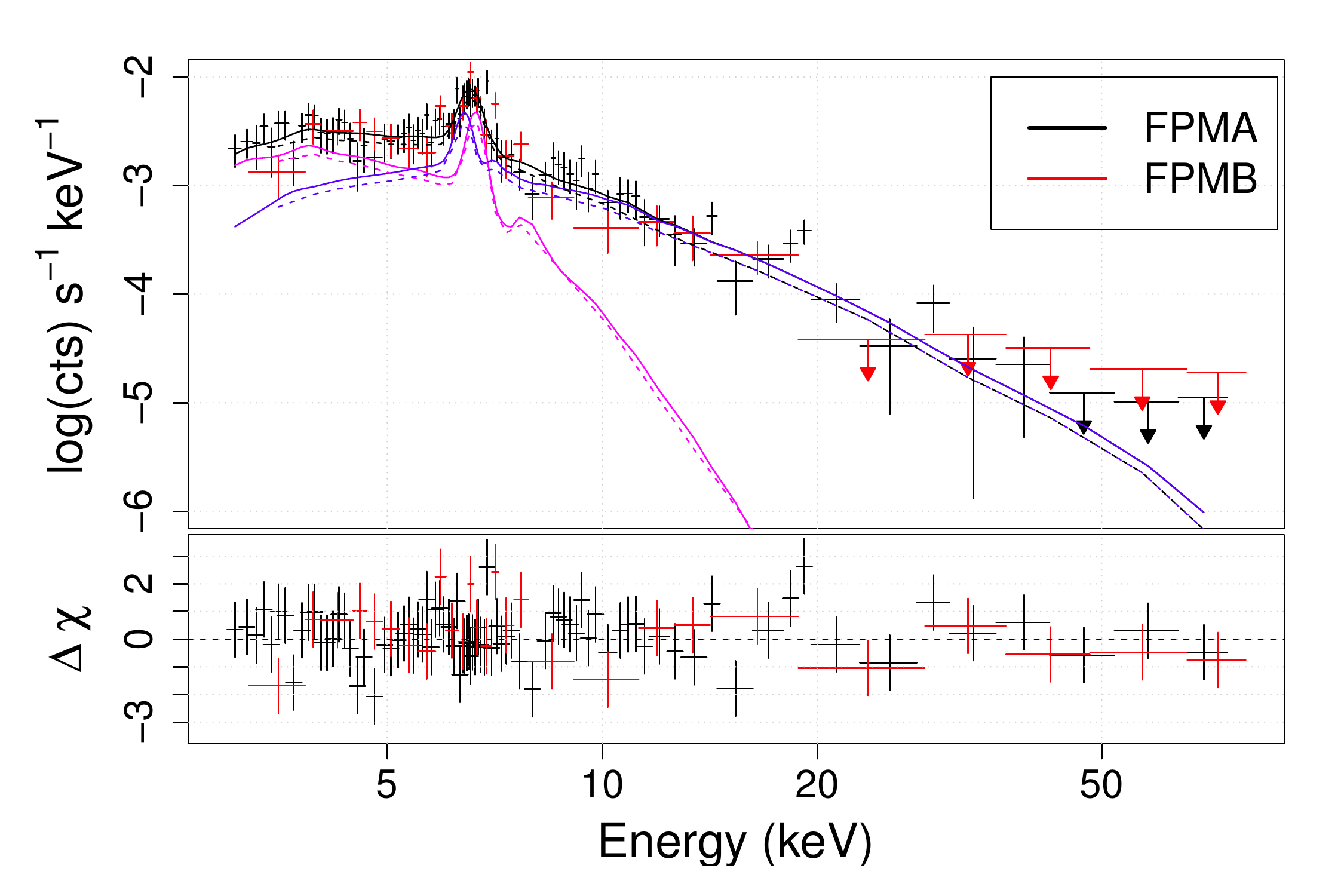}
\includegraphics[width=0.34\textwidth]{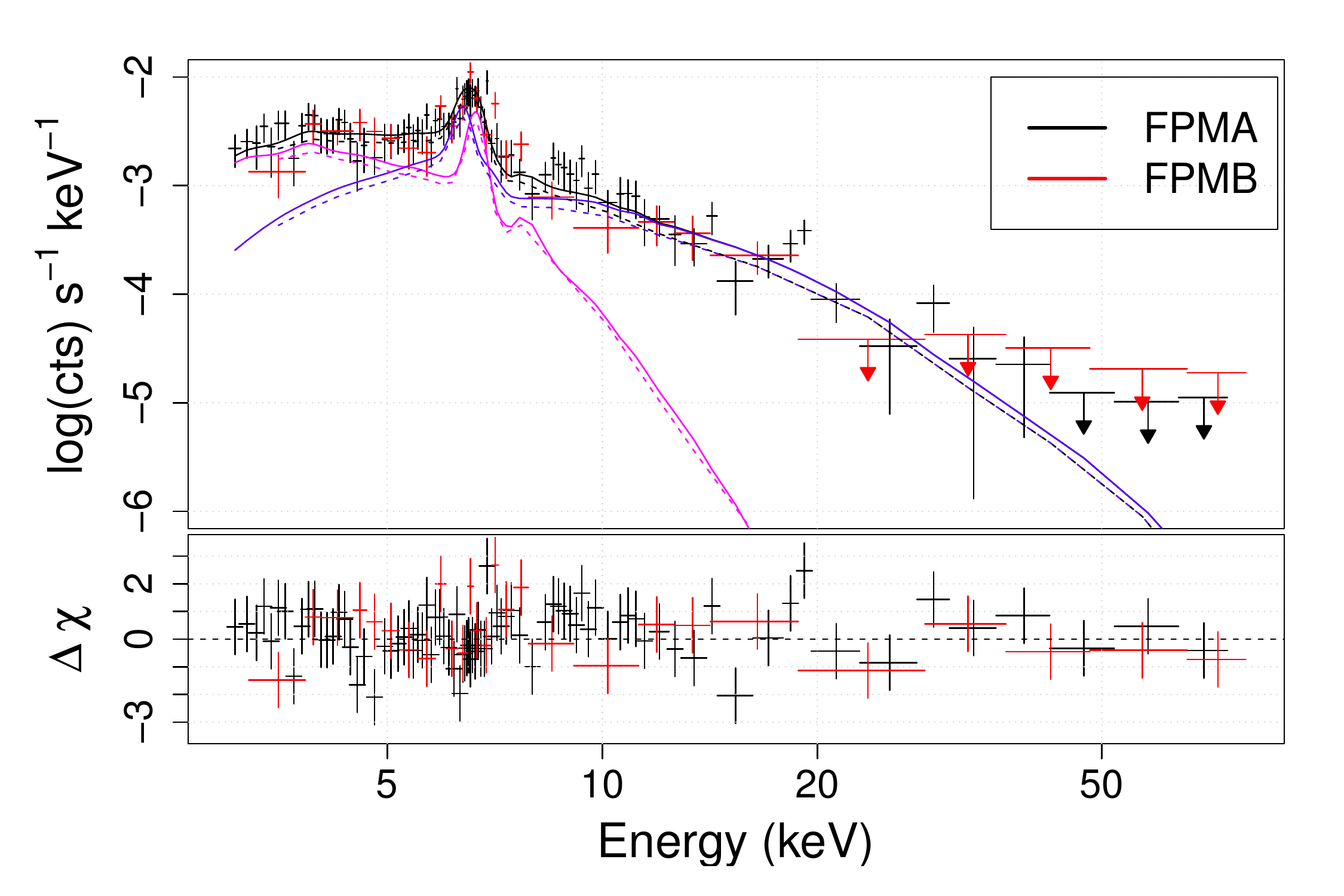}
\caption{The X-ray spectrum of the Arches cluster as measured with
  \textit{NuSTAR} FPMA (black) and FPMB (red). The spectra are fitted
  with models all containing {\small APEC} component (magenta) and: model~1
  (left) -- power-law (blue) and Gaussian (green); model~2 (middle) --
  {\small LECR}$p$ (blue); model~3 (right) -- {\small REFLIONX}
  (blue); see Table~\ref{tab:xspec} for details. Solid lines represent
  model components for FPMA, and dotted for FPMB, respectively.}
\label{fig:spec}
\end{figure*}

The extraction of the corresponding background spectra is complicated
by the strong CXB and GRXE stray-light illumination causing a
non-uniform pattern on the detectors with a gradient of a factor of
two. For that reason, the background spectrum extracted from the
source-free region over- or underpredicts the background count rate at
the position of the source. This difficulty (not relevant for spectra
of bright sources) causes strong systematic noise for the
background-subtracted spectrum of weak sources such as Arches. To
overcome this, we use the fact that the stray-light background does
not significantly change for sky pointings separated by only
$10-20'$, and therefore we utilize nearby observations. Note that this method
differs from that used in the imaging analysis. In the paired observation,
we extract the background spectrum from the same detector region where
the Arches cluster was observed.  Obviously, this region should
not contain any other diffuse or point-like source or stray-light from
nearby bright sources. Table~\ref{tab:pairs} lists the observations
used for source and background spectrum extraction. For \textit{ObsID}
40010006001, we had three observations which were suitable for
background spectrum extractions and thus averaged together for better
statistics.

Using the above-mentioned method, we have constructed one FPMB and
three FPMA spectra. The latter have been further grouped together using
the {\small FTOOL} task {\small MATHPHA}. The corresponding background
spectra and responses have been averaged accordingly. Finally we have
grouped spectrum channels to have at least 30 counts per bin using
{\small GRPPHA} tool.  Figure~\ref{fig:spec} shows the FPMA and FPMB
spectra of the Arches cluster.

According to the morphological studies presented in Sect.~\ref{morpho},
we attribute the hard spectral component above 10 keV to the
fluorescing cloud around the Arches cluster.

We first approximated the observed spectra with the collisionally
ionized plasma emission model {\small APEC} representing the cluster's
core thermal emission, and a non-thermal power-law continuum with a
Gaussian line at $6.4$~keV, modeling the cloud emission outside the
cluster (Figure~\ref{fig:spec}). Due to the fact that \textit{NuSTAR}
has a limited spectral resolution, the centroid energy of the Gaussian
line and its width have been fixed at $6.4$~keV and $0.1$~keV,
respectively.  All the emission components were subject to a
line-of-sight photoelectric absorption model {\small WABS} in {\small
  XSPEC}. The fitting procedure does not allow us to constrain the
absorption, and we fixed it at $N_{\mathrm{H}}=9.5\times 10^{22}$~cm$^{-2}$
measured by T12 (also using {\small WABS}) in the core of the Arches
cluster. The metallicity of the X-ray emitting plasma was not
constrained either, and we fixed it to $Z=1.7 Z_{\sun}$ as per
T12. The best-fit results obtained with this model (referred as
model~1 in the discussion) are shown in Table~\ref{tab:xspec}. All
fitted parameters in this table and in the following discussion
contain uncertainty estimation at the 90\% confidence level.

\begin{table}[htbp]
\begin{center}
  \caption{Observations used for source and
    background spectral extraction. \label{tab:pairs}}
\begin{tabular}{llcccc}
\tableline\tableline
\multicolumn{1}{c}{Source spectrum} & \multicolumn{1}{c}{Background spectrum}\\
\multicolumn{1}{c}{(ObsID$^a$)} & \multicolumn{1}{c}{(ObsID$^a$)} \\
\tableline
40010003001A$^b$  & 40010001001A \\
\tableline
40010005001A  & 40010003001A$^b$ \\
\tableline
40010006001A/B$^c$  & 40010001002A/B \\
                        & 40010003001A/B \\
                        & 40010005001A/B \\
\tableline
\end{tabular}
\end{center}
Notes. 
$^a$ The name of the focal plane module (A or B) is attached to the
end of the ObsID. 
$^b$ The corresponding FPMB data has strong stray-light contamination, see Sect.~\ref{data}. 
$^c$ The background spectrum has been averaged over
three observations.
\end{table}

Model~1 provides a reliable fit to the data by constraining the
temperature of the core emission, the power-law component and the
intensity of the $6.4$~keV line. The $1.76^{+0.36}_{-0.29}$~keV
temperature of the thermal component is in agreement with that
recently measured by T12 in the cluster's core
($1.79^{+0.06}_{-0.05}$~keV from ``model 1'' in T12). The power-law
slope $\Gamma=1.62\pm0.31$ is consistent with
$\Gamma=1.6^{+0.3}_{-0.2}$ from T12, measured in the cloud region. T12
finds the slope and $6.4$~keV line equivalent width
($EW_{\textrm{\tiny 6.4 keV}}$) of the cloud region compatible with
the low-energy cosmic-ray (LECR) ion model, where collisional
ionization of the cloud by cosmic ray (CR) protons produces the
$6.4$~keV fluorescence line. According to the \textit{XMM-Newton}
study of the Arches cluster by T12, the LECR {\it electron} model
(LECR$e$) requires an ambient Fe abundance $\gtrsim 3$ times the solar
value to account for the measured $EW$ of $1.2\pm0.2$~keV, which makes
the measured properties of the non-thermal cloud region emission
hardly compatible with the LECR electron model. We consider only the
LECR proton (LECR$p$) model in the following analysis.

\begin{table*}[htbp]
\begin{center}
  \caption{Spectral analysis of the Arches cluster X-ray emission. \label{tab:xspec}}
\begin{tabular}{cccccc}
\tableline\tableline
Parameter & (Unit) & Model 1 & Model 2 & Model 3 \\
\tableline
$C$ &  & $1.06\pm0.14$ & $1.05\pm0.14$ & $1.07\pm0.14$ \\
\tableline
$N_{\mathrm{H}}$ & $(10^{22} \textrm{cm}^{-2})$ & 9.5 (fixed) & 9.5 (fixed) & 9.5 (fixed) \\
$Z/Z_\sun$ &  & 1.7 (fixed) & 1.7 (fixed) & 1.7 (fixed) \\
\tableline
$kT$ & (keV) & $1.76^{+0.36}_{-0.29}$ & $1.69^{+0.31}_{-0.26}$ & $1.64^{+0.37}_{-0.23}$ \\
I$_{kT}$ & (see notes) & $36.0^{+17.8}_{-12.9}$ & $39.6^{+16.7}_{-11.9}$ & $43.2^{+16.9}_{-15.0}$ \\
\tableline
E$_{\textrm{\tiny 6.4 keV}}$ & (keV) & 6.4 (fixed) & -- & -- \\
F$_{\textrm{\tiny 6.4 keV}}$ & ($10^{-5}\textrm{ph cm}^{-2}\textrm{s}^{-1}$)
& $1.24\pm0.5$ & -- & -- \\
\tableline
$\Gamma$ &  & $1.62\pm 0.31$  & -- & -- \\
I$_{\textrm{\tiny p.l.}}$ & ($10^{-12}$erg~cm$^{-2}$s$^{-1}$ over 3-20~keV) & $1.49^{+0.26}_{-0.24}$ & -- & -- \\
\tableline
EW$_{\textrm{\tiny 6.4 keV}}$ & (keV) & $1.1^{+0.7}_{-0.5}$ & -- & -- \\
\tableline
$\Lambda$ & (H-atoms cm$^{-2}$) & -- & $5\times10^{24}$ (fixed) & -- \\
$s$ &   & -- & $1.65^{+0.59}_{-0.55}$ & -- \\
$E_{\textrm{\tiny min}}$ & (keV/n)  & -- & $10^{4}$ (fixed) & -- \\
$N_{\textrm{\tiny LECR}}$ & ($10^{-8}\textrm{~erg~cm}^{-2}\textrm{~s}^{-1}$)  & -- & $8.08^{+1.82}_{-1.62}$ & -- \\
\tableline
$\Gamma_{\textrm{\tiny RX}}$ &  & -- & -- & $2.93\pm 0.31$ \\
$\xi$ & erg~cm~s$^{-1}$  & -- & -- & $10$ (fixed) \\
$I_{\textrm{\tiny RX}}$ & ($10^{-5}$) & -- & -- & $6.09\pm 5.00$ \\

\tableline
Reduced $\chi^2$ (d.o.f.) &  & 0.91 (134) & 0.89 (136) & 0.91 (140) \\
\tableline
\end{tabular}
\end{center}
Notes. Model~1:
$C\times\textrm{WABS}\times(\textrm{APEC}+\textrm{Gaussian}+\textrm{powerlaw})$; Model~2:
$C\times\textrm{WABS}\times(\textrm{APEC}+\textrm{LECR}p)$; 
Model~3:
$C\times\textrm{WABS}\times(\textrm{APEC}+\textrm{REFLIONX})$. The
cross-callibration constant term $C$ is fixed at unity for FPMA data and fitted for FPMB. 
The absorption column density is noted as $N_{\mathrm{H}}$. APEC thermal plasma
model is characterized by $kT$, Z/Z$_\sun$, and I$_{kT}$ parameters,
respectively, describing temperature, metallicity relative to solar,
and normalization in units of $10^{-18}\int n_{e}n_{H}dV/(4\pi
D^{2})$, where $n_e$ and $n_H$ are the electron and proton number
densities in units of $\textrm{cm}^{-3}$, and D is the distance in
cm. E$_{\textrm{\tiny 6.4 keV}}$  and F$_{\textrm{\textrm{\tiny 6.4 keV}}}$: centroid energy and flux
of the neutral or low-ionization Fe~K$\alpha$ line. $\Gamma$
and I$_{\textrm{\tiny p.l.}}$: index and $3-20$~keV pegged normalization of the
power-law component (model {\small PEGPWRLW}). $EW_{\textrm{\tiny 6.4 keV}}$: $EW$ of the $6.4$~keV line
with respect to the power-law continuum. $\Lambda$, $s$,
$E_{\textrm{\tiny min}}$, and
$N_{\textrm{\tiny LECR}}$: LECR ions path length, source spectrum index, minimum
energy, and model normalization. By definition
$\textrm{d}W/\textrm{d}t=4\pi D^{2}N_{\textrm{\tiny LECR}}$ is the power
injected in the interaction region by primary CR protons with energies
between $E_{\textrm{\tiny min}}$ and $E_{\textrm{\tiny max}}=1$~GeV, where D is the
distance to the source. $\Gamma_{\textrm{\tiny RX}}$, $\xi$, and
$I_{\textrm{\tiny RX}}$: the
power-law slope of the incident radiation, ionization parameter
(defined as $\xi = 4\pi F_{tot}/n_{H}$ where $F_{\textrm{\tiny tot}}$ is the total illuminating flux, and
$n_{H}$ is the density of the reflector), and normalization of {\small
  REFLIONX} model \citep{reflionx}. The metallicity parameter $Z$ for {\small LECR} and {\small
  REFLIONX} models is fixed at 1.7$Z_{\sun}$. Reduced $\chi^{2}$ (d.o.f.):
reduced $\chi^{2}$ and degrees of freedom.
\end{table*}

We fitted the spectra with the XSPEC model
$\textrm{WABS}\times(\textrm{APEC}+\textrm{LECR}p)$, henceforth
``model 2''. The model depends on the LECR path length in the ambient
medium, $\Lambda$, the minimum energy of the CR particles entering the
cloud, $E_{min}$, the power-law index of the CR source energy
spectrum, $s$, and the metallicity of the X-ray emission region,
$Z$. We fixed $\Lambda$ and $E_{\textrm{\tiny min}}$ parameters of the LECR$p$ model
according to T12, making free slope $s$ and normalization
$N_{\textrm{\tiny LECR}}$. The best-fit results
obtained with model~2 are given in Table~\ref{tab:xspec} and the
spectral components are shown in Figure~\ref{fig:spec}.


The \textit{NuSTAR}-measured spectral index of the injected low-energy
protons $s=1.65^{+0.59}_{-0.55}$ is in general agreement with the T12
value ($1.9^{+0.5}_{-0.6}$). The best-fit LECR model normalization for
the fixed $E_{\textrm{\tiny min}}=10$~MeV~nucleon$^{-1}$
$N_{\textrm{\tiny LECR}}=(8.08^{+1.82}_{-1.62})\times
10^{-8}\textrm{~erg~cm}^{-2}\textrm{~s}^{-1}$ is somewhat higher than
that measured by T12 ($N_{\textrm{\tiny
    LECR}}=(5.6^{+0.7}_{-0.3})\times
10^{-8}\textrm{~erg~cm}^{-2}\textrm{~s}^{-1}$). The corresponding
power injected by LECR protons lies in the range $(5-8)\times
10^{38}$~erg~s$^{-1}$.

To test the X-ray reflection scenario of the Arches cluster
non-thermal emission, we applied the self-consistent reflection model
{\small
  REFLIONX}\footnote{http://heasarc.gsfc.nasa.gov/xanadu/xspec/models/reflion.html}
in {\small XSPEC} describing the reflected spectrum for an optically
thick atmosphere of constant density, illuminated by radiation with a
power-law spectrum \citep{reflionx}, henceforth ``model~3''. The
{\small REFLIONX} spectrum contains modeled fluorescence lines and
continuum emission (Figure~\ref{fig:spec}). Developed for the surface
of hot accretion disks, {\small REFLIONX} can still be applied for
cold material around the Arches cluster, fixing the ionization
parameter $\xi$ at the lowest allowed value of $10$~erg~cm~s$^{-1}$. The
best-fit parameters of model~3 are shown in Table~\ref{tab:xspec}.


\section{Discussion}
\label{discussion}

In this section we discuss the results of this work in the context of
two scenarios proposed to explain the excitation mechanism of the
fluorescent emission around the Arches cluster.

\subsection{CR bombardment hypothesis}

A particle bombardment excitation scenario for the Fe fluorescence line
emission involves the interaction of CR ions with cold molecular
gas. T12 developed a steady-state slab model in which accelerated
particles penetrate a cloud of neutral gas at a constant rate until
they either stop or escape from the cloud. The model describes the
properties of the neutral Fe~K$\alpha$ line emission excited by
impacts of LECR ions. These authors argue that fluorescent $6.4$~keV
line emission of the Arches cluster is excited by LECR protons with a
relatively soft incident spectrum $s=1.9^{+0.5}_{-0.6}$. Our
\textit{NuSTAR} spectral analysis (model~2) implies a somewhat harder slope
$s=1.65^{+0.59}_{-0.55}$, however, both measurements are consistent
within the error intervals. T12 notes that production of $6.4$~keV
line photons by LECR collisional ionization is relatively inefficient
and that high power in LECRs is needed to produce the observed
fluorescent Fe~K$\alpha$ line flux.  As the authors suggest, a large
flux of LECR ions could be produced in the ongoing supersonic
collision between the star cluster and nearby molecular cloud.

Note that LECR$p$ model provides an equally good fit as the reflection
model discussed below. 

\subsection{X-ray reflection hypothesis}

By definition, the excitation of the neutral material of the cloud by
hard X-ray irradiation implies a transient or persistent X-ray source
outside or within the cloud. 

Assuming that the irradiation source is located in the Arches cluster,
the source luminosity required to produce the observed 6.4~keV line
flux can be estimated from the fluorescent model described in
\cite{sunyaev1998}. Applying this approach on the \textit{XMM-Newton}
data, T12 showed that the luminosity of the source in the cluster must
be at least $10^{36}$~erg~s$^{-1}$. However, the persistent X-ray
luminosity is far below this level. \cite{capelli11a} recently
reported X-ray flaring activity in the core of the Arches cluster,
probably originating from stellar wind interactions in one or more
massive binary systems. Such transient activity, in principle, can
trigger fluorescent emission of the neutral material outside the
Arches cluster, but the peak luminosity of the observed flare is far
below the threshold mentioned above.

The illuminating source could also be a transient X-ray binary
system. Their typical X-ray luminosity can reach levels on the order
of $10^{36}-10^{38}$~erg~s$^{-1}$ for high- and low-mass systems. If
the bright state of an X-ray binary lasted for more than 8~years at or
above $10^{37}$~erg~s$^{-1}$, it would be sufficient for the observed
$6.4$~keV flux \citep{capelli11b}. However, such a long outburst is
much longer than typical flaring timescales of X-ray transients
\citep{degenaar2010}, though, at least one exceptional case is known,
the neutron star X-ray transient KS~1731-260
\citep{sunyaev1989,sunyaev1990} which became quiescent after more
than a decade in the bright state \citep{wijnands2001,revnivtsev2002}.


The \textit{NuSTAR}-measured $EW$ of the $6.4$~keV line with respect
to the power-law continuum, $EW_{\textrm{\tiny 6.4 keV}}=1.1^{+0.7}_{-0.5}$~keV, is in agreement with previous
studies performed on spatially resolved fluorescent clouds around the
Arches cluster \citep[][T12]{capelli11b}. In the context of an X-ray
reflection scenario, the relatively high $EW$ of the Fe~K$\alpha$ line
indicates a scattered emission component (assuming normal Fe
abundance) or strong photoabsorption \citep{sunyaev1998}, which can
explain the fact that the direct source of photoionization is not seen
by the observer. However, according to \textit{NuSTAR} maps of the
Arches cluster above $10$~keV (Figure~\ref{map:mosaics}), for which the
efficiency of photoabsorption rapidly falls, no significant source has
been detected.


The Arches cluster spectrum is well-fitted with a model composed of 
an optically thin thermal plasma and an X-ray reflection component
(Figure~\ref{fig:spec}). The spectrum of the illuminating X-ray
source is estimated to have relatively soft slope
$\Gamma_{\textrm{\tiny RX}}=2.93\pm0.31$. We may assess the luminosity of the primary
source $L_{\textrm{\tiny RX}}$ required by the reflection component following the
method used by \cite{reis2013} to estimate the reflection fraction
$R$. This parameter is defined as a measure of the ratio of the
reflected flux to the continuum power-law flux. {\small REFLIONX}
calculates $R$ by dividing the extrapolated ($1\textrm{ eV}-1000$~keV)
{\small REFLIONX} flux by the $100\textrm{ eV}-1000$~keV power-law flux. On the
other hand, the reflection scaling factor $R$ is in fact the solid angle
$R=\Omega/2\pi$ subtended by the reflecting cloud ($R=0$ means no
reflection). We estimated the unabsorbed $0.3-8$~keV luminosity
$L_{\textrm{\tiny RX}}$ (at 8~kpc) for the tabulated $R$ values -- $0.5$, $1.0$
(source above a plane), $1.5$ and $2.0$ (source inside the cloud) as
$2.5$, $1.3$, $0.8$ and $0.6\times 10^{36}$~erg~s$^{-1}$,
respectively. \cite{wang2006} reported the $0.3-8$~keV luminosity of
the bright Arches cluster sources A1N/S and A2, respectively, as $7.2$
(A1N), $11$ (A1S), and $4.6\times 10^{33}$~erg~s$^{-1}$, which is
$1-2$ orders of magnitude below the luminosity of the illuminating
continuum required by the reflected component. This constraint is
consistent with what can be drawn from the observed $6.4$~keV line
flux, using the general idea of \cite{sunyaev1998}, as demonstrated
by T12.

The interpretation of the observed spectra with the X-ray reflection
model confirms that the Arches cluster itself is unlikely to be the
source of the cloud fluorescence emission.

The non-detection of Fe~K$\alpha$ line flux variability over more
than eight years of \textit{XMM-Newton} observations
\citep[][T12]{capelli11b} puts strong limits on the photoionization
excitation of molecular clouds around the Arches cluster by a hard
X-ray source. Nevertheless, the plausibility of the scenario with an
earlier outburst of Sgr~A* is under debate. The large scale spatial
distribution and time variation of the Fe-K fluorescence seen in the
GC region reveal the apparent superluminal propagation of a light
front illuminating a molecular nebula \citep{ponti2010}. In this
context, it is quite reasonable to anticipate X-ray photoionization of
the molecular clouds near the Arches cluster by Sgr~A*. A possible
decrease in the 6.4~keV flux with time would be expected in this
scenario. For instance, strong variability of the 6.4~keV emission has
been detected in Sgr~B2 \citep{terrier2010} after ten years of a
persistent state as shown by \cite{revnivtsev2004a} -- however see
\cite{inui2009}. \cite{capelli11b} found the relatively constant
$6.4$~keV flux of the Arches cluster clouds consistent with a long
outburst from Sgr~A* at a constant luminosity of $1.4\times
10^{39}$~erg~s$^{-1}$ \citep{ponti2010}. On the other hand, \cite{clavel2013} possibly
revealed a highly variable active phase of Sgr~A* during the past few
hundred years.

The measured power-law slope of the illuminating source in the model~3
(Sect.~\ref{spectra}) is softer ($\Gamma\approx 3$) than observed for
Sgr~A* flares \citep[][Barriere et al. 2013, submitted to
ApJ]{porquet2003b,porquet2008,nowak2012}.  Additionally, the slope is
softer than that estimated by \cite{revnivtsev2004b} for Sgr~B2 in the
context of Sgr~A* reflected emission. However, we should note that
{\small REFLIONX} calculates the reflected spectrum for an optically
thick atmosphere while the molecular clouds close to the Arches
cluster exhibit transparency at $\tau<1$ \citep{capelli11b}. A more
realistic reflection model at lower $\tau$ values and different
source-cloud geometry settings can produce a harder spectrum of the
primary emission.

\section{Summary}
\label{conclusion}

In this paper we present results on the first focused hard X-ray
observation of the Arches cluster performed by \textit{NuSTAR}. We
report on significant detection of the hard X-ray emission around the
cluster with a spectrum that extends for the first time up to $\sim
40$~keV. The main results can be summarized as follows:

\begin{itemize}
\item The \textit{NuSTAR} mosaic image of the Arches cluster shows a core
  component and extended halo emission in the $3-10$~keV energy band. The
  centroid position of the core emission coincides with the bright sources
  A1N/S resolved with \textit{Chandra} in the Arches cluster. The halo emission
  around the cluster also follows the morphology observed with
  \textit{Chandra}. The centroid position of the halo emission does not
  coincide with the core and tends to be outside the Arches cluster.

\item Wavelet decomposition of sky mosaics does not show any
  significant ($5\sigma$) detection of point-like sources in the core of
  the Arches cluster above $10$~keV. The continuum $10-20$~keV
  emission is significantly detected well beyond the Arches cluster
  with a spatial morphology consistent with the Fe~K$\alpha$ fluorescent
  $6.4$~keV line emission observed with \textit{XMM-Newton} (T12).

\item The spectral analysis of the Arches cluster region, including
  the extended emission around it, reveals a $\sim1.7$~keV temperature
  plasma, which is consistent with the optically thin thermal plasma
  measured by T12 in the cluster's core and associated with
  thermalization of massive star winds and several colliding stellar
  wind binaries within the cluster (T12).

\item The power-law slope $\Gamma= 1.62\pm 0.31$ and
  $EW_{\textrm{\tiny 6.4 keV}}=1.1^{+0.7}_{-0.5}$~keV are in agreement with that
  measured by T12 around the Arches cluster and expected for the LECR ion
  model.

\item The X-ray emission of the Arches cluster is well fitted with a
  model composed of an optically thin thermal plasma and a non-thermal
  component produced by LECR ions (T12). The best-fit CR spectral
  index is $s=1.65^{+0.59}_{-0.55}$. The LECR model normalization
  gives the corresponding power of the injected LECR protons in the
  cloud region at the level of $(5-8)\times 10^{38}$~erg~s$^{-1}$
  (with D=8~kpc and CR minimum energy $E_{\textrm{min}}=10$~MeV~nucleon$^{-1}$).

\item The luminosity of the primary source required by the X-ray
  reflection model is $1-2$ orders of magnitude higher than that
  observed in the Arches cluster, which is in agreement with similar
  constraints drawn from the 6.4~keV line flux.
\end{itemize}

The X-ray photoionization and CR-induced emission models can reproduce
the data equally well. Additional \textit{NuSTAR} observations with
the Arches cluster on-axis would be useful to detect the difference
between the emission models above $20$~keV. At this point, we can
conclude that explaining the non-thermal emission around the Arches
cluster with an X-ray reflection scenario in which the illuminating
source is located in or close to the cluster is unlikely. However
irradiation by a long outburst of Sgr~A* can still be considered, and
would imply long-term variability of the 6.4~keV line flux should be
detected. The excitation of the neutral material around the cluster by
CR interactions also looks plausible, though a large flux of LECR
protons is needed to produce the observed fluorescent emission.


\acknowledgments

This work was supported under NASA Contract No. NNG08FD60C, and made
use of data from the \textit{NuSTAR} mission, a project led by the
California Institute of Technology, managed by the Jet Propulsion
Laboratory, and funded by the National Aeronautics and Space
Administration. We thank the \textit{NuSTAR} Operations, Software and
Calibration teams for support with the execution and analysis of these
observations. This research has made use of the \textit{NuSTAR} Data
Analysis Software (\textit{NuSTARDAS}) jointly developed by the ASI
Science Data Center (ASDC, Italy) and the California Institute of
Technology (USA). FEB acknowledges support from Basal-CATA
(PFB-06/2007) and CONICYT-Chile (FONDECYT 1101024 and Anillo
ACT1101). RK thanks Eugene Churazov for fruitful discussions and
valuable suggestions to the paper.




{\it Facilities:} \facility{NuSTAR}.





\end{document}